\documentclass[a4paper,12pt]{article}
\usepackage{amssymb,amsmath,cite,mathdots}
\usepackage{setspace}
\usepackage[T1]{fontenc}
\usepackage{times}
\usepackage{palatino} 
\usepackage{lmodern}
\usepackage[latin1]{inputenc}
\usepackage{epsfig}
\usepackage[english]{babel}
\usepackage{color}
\usepackage{graphicx}
\usepackage{dcolumn}
\usepackage{moreverb}
\usepackage{amsmath,amssymb,amsfonts}
\usepackage[all]{xy}
\usepackage{subfig}

\newtheorem{lemma}{Lemma}
\newtheorem{prop}{Proposition}

\newtheorem{cor}{Corollary}
\newtheorem{defo}{Definition}

\def\proof#1{{\bf Proof:} #1 $\blacksquare$ \medskip}

\def\nn{\nonumber}

\begin{document}
\numberwithin{equation}{section}
\newcommand{\boxedeqn}[1]{%
  \[\fbox{%
      \addtolength{\linewidth}{-2\fboxsep}%
      \addtolength{\linewidth}{-2\fboxrule}%
      \begin{minipage}{\linewidth}%
      \begin{equation}#1\end{equation}%
      \end{minipage}%
    }\]%
}

\begin{center}
{\large\bf On realizations of polynomial algebras with three generators via deformed oscillator algebras}\\
~~\\

{\large Phillip S. Isaac and Ian Marquette}\\
~~\\

School of Mathematics and Physics, The University of Queensland, St Lucia QLD 4072, Australia.
\end{center}

\begin{abstract}
 
We present the most general polynomial Lie algebra generated by a second order
integral of motion and one of order $M$,
construct the Casimir operator,
and show how the Jacobi identity
provides the existence of a realization in terms of deformed oscillator algebra. 
We also present the classical analog of this construction for the most general Polynomial Poisson
algebra. Two specific classes of such polynomial algebras are discussed that
include the symmetry algebras observed for various 2D superintegrable systems.
\end{abstract}

%
\section{Introduction}

The study of quadratically superintegrable systems using quadratic algebra was initiated
by Zhedanov et al. \cite{Zhe1} and since, various examples of quadratically superintegrable
systems have been studied using this approach
\cite{Zhe1,Vin1,Das1,Kre1,Pos1,Que1,Das2,Mar1,Gen1,Gen2,Mil1}. 
These quadratic algebras that are quadratic extensions of Lie algebras
appear to be very rich objects that allow one to obtain algebraically the energy spectrum of
superintegrable systems and explain the total number of degeneracies for a given level through
representation theory even beyond two dimensional cases \cite{Das2,Mar1}. 
These algebraic structures can also be used to classify superintegrable Hamiltonians \cite{Kre1}
using the Casimir operators.
In addition, there is a very interesting and recent connection between quadratic
algebras of superintegrable systems in two-dimensional conformally flat spaces with the
full Askey scheme of orthogonal polynomials \cite{Mil1}. For a contemporary review on the topic of
classical and quantum systems and related algebraic structures with, in addition, an extensive list of
references, we refer the reader to \cite{Mil2}.

There exist realizations of quadratic associative algebra in terms of deformed oscillator
algebra \cite{Das1}. A deformed oscillator algebra generated by $\{N,b,b^{\dagger},1\}$ has the
following form
\cite{Das3,Van1}:
\begin{equation*}
[N,b^{\dagger}]=b^{\dagger} ,\quad [N,b]=-b ,\quad b^{\dagger}b=\Phi(N) ,\quad
bb^{\dagger}=\Phi(N+1).
\end{equation*}
It was extended for specific cases to take into account reflection and grading \cite{Van2,Plyu} and many papers were devoted to deformed
oscillator beyond q-algebras and their applications \cite{Gav1,Gav2,Bur1,Gav3,Gaz1,Bal1}.

Initiated by Daskaloyannis for quadratic associative algebra related to superintegrable systems and generated by two second order integrals of motion, 
the approach of constructing realizations via deformed oscillator algebras 
can be used to obtain finite-dimensional unitary representations. This construction of Daskaloyannis has been
extended to cubic \cite{Mar2} and quartic \cite{Mar3} associative algebras in order to
study their representations and apply them to various superintegrable systems with higher
order integrals of motion. 
Polynomial algebras of arbitrary order have been obtained for specific examples of superintegrable quantum systems using various
approaches \cite{Mar4,Kal1} involving recurrence relations and ladder operators. However, the existence of realizations as deformed oscillator algebras for general polynomial Lie 
algebras beyond the quartic case is an unexplored subject.
The purpose of this paper is to study the existence of a Casimir operator and realizations via deformed oscillator
algebras for polynomial algebras of arbitrary order with three generators 
and the connection with the Jacobi identity.

Let us describe the organization of the paper. In Section \ref{Section2}, we present the most general polynomial
Poisson algebra with three generators associated with a superintegrable systems with two integrals
of motion of order 2 and $M$.
We present constraints from the Jacobi identity and calculate the Casimir operator. We obtain
realizations in terms of the classical analogue of the deformed
oscillator algebra. In Section \ref{Section3}, we study the corresponding polynomial Lie algebras with three
generators and using the Jacobi identity obtain constraints on the structure constants and for the existence of a Casimir
operator. We present an algorithm to solve these constraints at any order $M$ from linear recurrence equations.
We show that the Jacobi identity provides the existence of a realization via deformed oscillator
algebras and present the Casimir operator in terms of the parafermionic number only. 
In Section \ref{Section4}, we discuss more explicitly two families of polynomial algebras of arbitrary order
that are particularly relevant to superintegrable systems.

\section{Polynomial Poisson algebras with 3 generators} \label{Section2}

Here we begin with the classical case and consider only two dimensional systems with
canonical position $q_1,$ $q_2$ and canonical momenta $p_1,$ $p_2$. 
We seek to introduce the most general polynomial Poisson algebra
${\cal P}_M$ equipped with Poisson bracket denoted $\{,\}_p,$ and generated by a function
$A$ which is second order polynomial in the momenta and another integral $B$ of
arbitrary order $M$. That is, 
\begin{equation}
A=\sum_{0\leq i+j \leq 2}f_{ij}(q_{1},q_{2})p_{1}^{i}p_{2}^{j},
\end{equation}
\begin{equation}
B=\sum_{0\leq i+j \leq M}g_{ij}(q_{1},q_{2})p_{1}^{i}p_{2}^{j},
\end{equation}
where the $f_{ij}$ and $g_{ij}$ are sufficiently smooth functions of the canonical position
coordinates only.
Note that on all suitably smooth functions of the canonical coordinates, the Poisson
bracket has the well known form
\begin{equation}
\{X,Y\}_p = 
\sum_{i=1}^2\left( \frac{\partial X}{\partial q_i}\frac{\partial Y}{\partial p_i} 
 - \frac{\partial X}{\partial p_i}\frac{\partial Y}{\partial q_i}
\right).
\label{PoissonBracketForm}
\end{equation}
We note that the Poisson bracket is antisymmetric, i.e. $\{X,Y\}_p = -\{ Y,X\}_p$, and
the Jacobi identity satisfied by $\{,\}_p$ is in general given by
\begin{equation}
\{X,\{Y,Z\}_p\}_p+\{Y,\{Z,X\}_p\}_p+\{Z,\{X,Y\}_p\}_p=0,
\label{generaljac}
\end{equation}
for all relevant functions $X,Y,Z$. We also remark that the Poisson bracket
satisfies the derivation rule,
\begin{equation}
\{X,YZ\}_p = \{X,Y\}_p Z + \{ X,Z\}_p Y.
\label{derivation}
\end{equation}

\subsection{Explicit Poisson bracket}

For convenience, we define a third function $C$ by 
$$
C = \{A,B\}_{p},
$$
and observe from (\ref{PoissonBracketForm}) that it is of order $M+1$ in the momenta.

In the case under consideration, the only non-trivial expression for the Jacobi identity
in equation (\ref{generaljac}) that we need to consider is
\begin{equation}
\{A,\{B,C\}_{p}\}_{p}=\{B,\{A,C\}_{p}\}_{p}.
\label{jacobi}
\end{equation}
This leads to the following result.

\begin{prop} \label{classicalrels}
The polynomial Poisson algebra ${\cal P}_M$ generated by functions $A$ and $B$ of
order 2 and $M$ respectively in the momenta, has Poisson bracket given by 
\begin{subequations}
\begin{equation}
\{A,B\}_{p}=C, \label{genl1v2c}
\end{equation}
\begin{equation}
\{A,C\}_{p}=\sum_{i=1}^{\left\lfloor\frac{M}{2}+1\right\rfloor}\alpha_{i}A^{i}+\delta B +\epsilon +2 \beta AB, \label{genl2v2c}
\end{equation}
\begin{equation}
\{B,C\}_{p}=\sum_{i=1}^{M}\lambda_{i}A^{i}-\beta B^{2}
-\alpha_{1}B-\sum_{i=1}^{\left\lfloor\frac{M}{2}\right\rfloor}(i+1)\alpha_{i+1}A^{i}B+\zeta. \label{genl3v2c}
\end{equation}
\end{subequations}
\end{prop}

\proof{
We begin with arbitrary forms of appropriate order for the brackets $\{A,C\}_p$ and $\{B,C\}_p$, namely
\begin{subequations}
\begin{equation}
 \{A,C\}_{p}=\sum_{i=1}^{\left\lfloor\frac{M}{2}+1\right\rfloor}\alpha_{i}A^{i}+\delta B +\epsilon +2\beta A B,  \label{genl2v1c}
\end{equation}
\begin{equation}
 \{B,C\}_{p}= \sum_{i=1}^{M}\lambda_{i}A^{i}+\rho B^{2} +\eta B +
\sum_{i=1}^{\left\lfloor\frac{M}{2}\right\rfloor}2\omega_{i} A^{i} B +\zeta, \label{genl3v1c}
\end{equation}
\end{subequations}
where we use the standard notation $\lfloor y\rfloor$ to denote the integer part of $y$.
The forms
of the right hand sides of equations~\eqref{genl2v1c} and~\eqref{genl3v1c} are determined
by allowing the most general polynomial in generators $A$ and $B$ constrained by the order 
of the Poisson bracket on the left side. These expressions do not depend on the generator $C$.
The Jacobi identity (\ref{jacobi}) provides constraints on the structure constants of the polynomial Poisson algebra,
which are shown to be given by the linear relations
\begin{equation}
\eta=-\alpha_{1},\quad \rho=-\beta,\quad 2\omega_{i}=-(i+1)\alpha_{i+1}, \label{Jaccon}
\end{equation}
thus proving the form of the Poisson algebra as stated.
}

\subsection{Casimir operator} \label{casimiropsec}

Let $K$ denote the Casimir operator, defined by
\begin{equation}
\{K,A\}_p =0 = \{K,B\}_p
\label{CasimirDef}
\end{equation}
in this case.
For polynomial Poisson algebras of low orders, the Casimir operator is known
\cite{Das1,Mar2,Mar3} to have the form 
$$
K=C^{2}+P(A,B),
$$ 
where $P(A,B)$ is a polynomial of the same order as $C^{2}$ in terms of momenta. Consequently, the following
result is established immediately.
\begin{prop} \label{casimirthm}
The Casimir operator $K$ for ${\cal P}_M$ is given by
$$
K=C^{2}-\sum_{i=1}^{\left\lfloor\frac{M}{2}+1\right\rfloor}2\alpha_{i}A^{i}B-2\beta
AB^{2}+2\zeta A + \sum_{i=1}^{M}\frac{2}{i+1}\lambda_iA^{i+1}-2\epsilon B - \delta B^2.
$$
\end{prop}
\proof{
The most general form of the Casimir operator with terms up to the same order as $C^2$ is given by
$$
K=C^{2}+\sum_{i=1}^{\left\lfloor\frac{M}{2}+1\right\rfloor}m_{i}A^{i}B+n AB^{2}+\sum_{i=1}^{M+1}k_{i}A^{i}+\sum_{i=1}^{2}l_{i}B^{i}.
$$
Substituting this form into the definition given in equation (\ref{CasimirDef}), the
parameters $m_{i}$, $n$, $k_{i}$ and $l_{i}$ are found to be expressible in terms
of the structure constants as
\begin{equation}
m_{i}=-2\alpha_{i},\quad l_{1}=-2\epsilon,\quad l_{2}=-\delta,\quad n=-2\beta,  \label{Cascon}
\end{equation}
\[ k_{1}=2\zeta,\quad k_{i+1}=\frac{2}{i+1}\lambda_{i}. \]
}

Following \cite{GMP}, we make the remark that by setting $F=-\frac{1}{2}P(A,B),$
the Poisson bracket is expressible in terms of $F$ as 
$$
\{A,B\}_{p}=C,\quad \{A,C\}_{p}=\frac{\partial F}{\partial B},\quad \{B,C\}_{p}=-\frac{\partial F}{\partial A}.
$$

\subsection{Oscillator realization}

Let us now investigate realizations of ${\cal P}_M$ in terms of the classical analogue of
the deformed oscillator algebra $\{1,N,b^{+},b\}$ with relations
\begin{equation}
\{N,b\}_p=-b,\quad \{N,b^{+}\}_p=b^{+},\quad
bb^{+}=b^{+}b=G(N),\quad \{b,b^{+}\}_{p}=\Phi(N),
\label{classicalrealisation}
\end{equation}
where $G(N)$ and $\Phi(N)$ are as yet undetermined functions. Let us note that the first and 
second equation of (\ref{classicalrealisation}) provide the third one using the Jacobi identity
and the fourth one using the derivation rule. Furthermore, it can be shown that

\begin{equation}
\Phi(N)=G'(N).
\label{relationphig}
\end{equation}

We start with a straightforward lemma that will be useful in our calculations.

\begin{lemma} \label{lem1}
Let $x(N)$ be a function of $N$ expressible as a formal power series. Then
\begin{align}
\{ x(N),b\}_p &= -x'(N)b,\nn\\
\{ x(N),b^+\}_p &= x'(N)b^+,\nn
\end{align}
where $x'(N)$ denotes the usual derivative of $x(N)$.
\end{lemma}
\proof{
Using induction and the derivation rule (\ref{derivation}), it is a straightforward matter to establish the relations
\begin{align}
\{ N^k,b\}_p &= -kN^{k-1}b,\nn\\
\{ N^k,b^+\}_p &= kN^{k-1}b^+,\nn
\end{align}
for any $0\leq k\in\mathbb{Z}$.
The result is then obtained by expressing $x(N)$ as a formal power series.
}

Our goal is to consider realizations of the form
\begin{equation*}
A=A(N),\quad B=b(N)+\rho(N)b + \rho(N)b^{+} ,
\end{equation*}
and to then determine constraints on the functions $A(N)$, $b(N)$ and $\rho(N)$,
along with $G(N)$ and $\Phi(N)$. To derive such constraints, we begin by imposing the
relations of Theorem \ref{classicalrels}, and make repeated use of the result of Lemma
\ref{lem1}.

Firstly, equation~(\ref{genl1v2c}) gives 
\begin{equation*}
\{A(N),b(N)+\rho(N)b + \rho(N)b^{+}\}_{p}=C,
\end{equation*}
from which we obtain
\begin{equation*}
C=\rho(N)A(N)'(b^{+}-b).
\end{equation*}

Equation~\eqref{genl2v2c} then gives 
\begin{align*}
\{ A(N),\rho(N)A'(N)(b^+-b)\}_p &= \sum_{i=1}^{\left\lfloor
\frac{M}{2}+1\right\rfloor}\alpha_i A(N)^i + \delta(b(N) + \rho(N)b+\rho(N)b^+) + \epsilon
\\
& \qquad \qquad + 2\beta A(N)(b(N) + \rho(N)b+\rho(N)b^+)\\
\Rightarrow \ \ \rho(N) A'(N)^2(b+b^+) &= 
\sum_{i=1}^{\left\lfloor \frac{M}{2}+1\right\rfloor}\alpha_i A(N)^i + \delta b(N) + \epsilon
+ 2\beta A(N)b(N)\\
& \qquad\qquad + (\delta + 2\beta A(N))\rho(N)(b+b^+).
\end{align*}
By equating coefficients of $(b+b^+)$ and the remaining functions of $N$, assuming
$\rho(N)\neq 0$, we obtain the following two
constraints on the unknown functions $A(N)$ and $b(N)$.
\begin{align}
& A'(N)^{2}=\delta+2\beta A(N), \label{solcAB1} \\
& \sum_{i=1}^{\left\lfloor\frac{M}{2}+1\right\rfloor}\alpha_{i}A(N)^i+\delta b(N) +2 \beta
A(N) b(N)+\epsilon=0. \label{solcAB2}
\end{align}
Assuming that $A(N)$ is non-trivial, particularly that $A'(N)\neq0$, we
note that differentiating these two constraints respectively gives
\begin{align}
& 2A'(N)A''(N)=2\beta A'(N) \ \ \Rightarrow \ \  A''(N) = \beta, \label{solcAB1diff} \\
& \sum_{i=1}^{\left\lfloor\frac{M}{2}+1\right\rfloor}i\alpha_{i}A'(N)A(N)^{i-1}+\delta
b'(N) +2 \beta( A'(N) b(N)+ A(N)b'(N))=0. \label{solcAB2diff}
\end{align}

Finally, equation~\eqref{genl3v2c} gives rise to three identities, obtained by
equating the coefficients of $(b^2+\left( b^+ \right)^2)$, $(b+b^+)$ and the remaining functions
of $N$. Assuming $\rho(N)\neq 0$, the coefficients of $(b^2+\left( b^+ \right)^2)$ and $(b+b^+)$
give, respectively, the identities
\begin{align}
&A''(N) = \beta \label{rel31},\\
& A'(N) b'(N) =
-2\beta b(N)-\alpha_1-\sum_{i=1}^{\left\lfloor\frac{M}{2}\right\rfloor}(i+1)\alpha_{i+1}A(N)^i.
\label{rel32}
\end{align}
Equations (\ref{rel31}) and (\ref{solcAB1diff}) are precisely the same. Moreover, if we
multiply equation (\ref{rel32}) by $A'(N)$ and impose the constraint (\ref{solcAB1}), we
may easily arrive at equation (\ref{solcAB2diff}). It is clear, then, that equations (\ref{rel31})
and (\ref{rel32}) give no new information. Equating the remaining
functions of $N$ in equation (\ref{genl3v2c}), however, gives the only constraint so far
involving the structure functions $G(N)$ and $\Phi(N)$,
namely
\begin{align}
& 4\rho'(N)\rho(N)A'(N)G(N)+2\rho(N)^{2}A''(N)G(N)+2\rho(N)^{2}A'(N)\Phi(N)\nn\\
& \qquad =\sum_{i=1}^{M}\lambda_{i}A(N)^{i}
-\beta
b(N)^{2}-2\beta\rho(N)^2G(N)-\alpha_{1}b(N)-b(N)\sum_{i=1}^{\left\lfloor\frac{M}{2}\right\rfloor}(i+1)\alpha_{i+1}
A(N)^{i}b(N)+\zeta.
\label{eq3c3}
\end{align}
Taking equation (\ref{rel32}) and multiplying through by $b(N)$ (which we assume to be
non-zero) gives
$$
A'(N) b'(N)b(N) + \beta b(N)^2 = -\beta b(N)^2-b(N)\alpha_1-b(N)\sum_{i=1}^{\left\lfloor
\frac{M}2\right\rfloor}(i+1)\alpha_{i+1}A(N)^i.
$$
We can use this, along with equations (\ref{rel31}) and (\ref{relationphig}), to further simplify equation (\ref{eq3c3}). 

The discussion of this section is summarised in the following, noting that the differential equation (\ref{solcAB1}) may be easily solved.
\begin{prop} \label{realthm}
The Poisson algebra ${\cal P}_M$ has the realization
\begin{align*}
A &= A(N),  \\
B &= b(N) + \rho(N)(b+b^+),\\
C &= \rho(N)A'(N)(b^+-b),
\end{align*}
in terms of the classical analogue of
the deformed oscillator algebra with relations given by (\ref{classicalrealisation}),
where
\begin{align*}
A(N) &= \left\{ \begin{array}{rl}
\displaystyle{\sqrt{\delta}N + c_1},& \beta=0,\\
\displaystyle{-\frac{\delta}{2\beta}+\frac{\beta}{2}(N+c_{1})^{2}},& \beta\neq 0,
\end{array} \right.
\end{align*}
with $c_1$ an arbitrary constant,
\begin{align*}
b(N) &=  -\frac{1}{A'(N)^2}\sum_{i=1}^{\left\lfloor\frac{M}{2}+1\right\rfloor}
\alpha_{i}A(N)^i+\epsilon,
\end{align*}
and the function $\rho(N)$ along with the structure functions $G(N)$ and $\Phi(N)$ must satisfy the constraint
\begin{align*}
& [A'(N)(2\rho(N)^{2}G(N))]'   \nn\\
& \qquad =\sum_{i=1}^{M}\lambda_{i}A(N)^{i}
+A'(N)b'(N)b(N) + \beta b(N)^2 +\zeta.
\end{align*}
\end{prop}

\begin{cor}
The Casimir operator $K$ for ${\cal P_M}$ has a realization in terms of the classical
analogue of the deformed oscillator algebra (satisfying relations (\ref{classicalrealisation})) given by
\begin{align*}
K &=  
-2\rho(N)^2A'(N)^2G(N)
-\sum_{i=1}^{\left\lfloor\frac{M}{2}+1\right\rfloor}2\alpha_{i}A(N)^ib(N)
-2\beta A(N) (b(N)^2+2 \rho(N)^2G(N))\\
& \qquad +2\zeta A(N)
\sum_{i=1}^{M}\frac{2}{i+1}\lambda_iA(N)^{i+1} -2\epsilon
b(N)-\delta (b(N)^2+2\rho(N)^2G(N)).
\end{align*}
\end{cor}
\proof{
The abstract form of the Casimir operator was determined in Proposition \ref{casimirthm}.
Substituting the realization from Proposition \ref{realthm},
\begin{align*}
A &= A(N),  \\
B &= b(N) + \rho(N)(b+b^+),\\
C &= \rho(N)A'(N)(b^+-b),
\end{align*}
into the expression from Proposition \ref{casimirthm} for $K$ gives 
\begin{align*}
K &= 
\rho(N)^2\left( A'(N)^2 - 2\beta A(N) -\delta \right) \left( b^2+\left(b^+\right)^2 \right) \\
& \qquad -2\rho(N)\left( \sum_{i=1}^{\left\lfloor\frac{M}{2}+1\right\rfloor}\alpha_{i}A(N)^i+\delta b(N) +2 \beta
A(N) b(N)+\epsilon \right) \left( b+b^+ \right) \\
& \qquad -2\rho(N)^2A'(N)^2G(N)
-\sum_{i=1}^{\left\lfloor\frac{M}{2}+1\right\rfloor}2\alpha_{i}A(N)^ib(N)
-2\beta A(N) (b(N)^2+2 \rho(N)^2G(N))\\
& \qquad +2\zeta A(N)
\sum_{i=1}^{M}\frac{2}{i+1}\lambda_iA(N)^{i+1} -2\epsilon
b(N)-\delta (b(N)^2+2\rho(N)^2G(N)).
\end{align*}
The coefficients of $\left( b^2+\left(b^+\right)^2 \right)$ and $\left( b+b^+ \right)$
are both zero due to the imposed constraints (\ref{solcAB1}) and (\ref{solcAB2}), hence the result.
}

We remark that the function $\rho(N)$ can be chosen so that the structure function
$\Phi(N)$ is a polynomial.


\section{Polynomial Lie algebras with three generators} \label{Section3}

Let us now consider the quantum case. We replace the Poisson bracket by the commutator,
and make use of the quantum operators $A$ and $B$, being analogues of their classical
counterparts. Quadratic terms such as $AB$ are then symmetrised using the anticommutator $\{A,B\}$.
The explicit calculations to any given order can be done but are much more involved even
for low order (i.e. small values of $M$). We will ultimately show that a realization
exists when we impose the Jacobi identity. 

\subsection{Explicit Lie bracket}

By generalisation of the classical case covered in the previous section, we define an
operator $C$ by
$$
C = [A,B],
$$
where $[A,B]=AB-BA$ denotes the commutator. To be a Lie algebra, the structure constants
must be defined so as to satisfy the Jacobi identity, which has the general form
$$
[X,[Y,Z]] - [Y,[X,Z]] = [[X,Y],Z],
$$
for operators $X,Y,Z$. For the case at hand, the only non-trivial situation required for
consideration is
$$
[A,[B,C]] = [B,[A,C]].
$$
To prove the following proposition, it is a simple matter of substituting the forms of the
given Lie brackets into this Jacobi identity. We also use the notation $\{ X,Y\} = XY+YX$
for the anticommutator.

\begin{prop} \label{prop4}
The polynomial Lie algebra, ${\cal L}_M$, which is the $M$th order analogue of the classical Poisson
algebra ${\cal P}_M$ of Proposition \ref{classicalrels}, has bracket operation given by
\begin{subequations}
\begin{equation}
[A,B]=C,     \label{genl1v1q}
\end{equation}
\begin{equation}
 [A,C]=\sum_{i=1}^{\left\lfloor\frac{M}{2}+1\right\rfloor}\alpha_{i}A^{i}+\delta B
+\epsilon + \beta \{A,B\},    \label{genl2v1q}
\end{equation}
\begin{equation}
[B,C]=\sum_{i=1}^{M}\lambda_{i}A^{i}-\beta B^{2} +\eta B +
\sum_{i=1}^{\left\lfloor\frac{M}{2}\right\rfloor}\omega_{i}\{A^{i},B\}+\zeta
\label{genl3v1q}
\end{equation}
\end{subequations}
subject to the constraint
\begin{equation}
\eta C+\sum_{i=1}^{\left\lfloor\frac{M}{2}\right\rfloor}\omega_{i}\{A^{i},C\}
+\sum_{k=1}^{\left\lfloor\frac{M}{2}+1\right\rfloor}\alpha_{k}[A^{k},B]=0. \label{qjaccon2}
\end{equation}
\end{prop}

The constraint in Proposition \ref{prop4} is difficult to evaluate in general. Here we
seek to develop an algorithm to determine conditions on the coefficients $\eta,$ $\omega_i$ and
$\alpha_i$ so that equation (\ref{qjaccon2}) is satisfied. To this end, the following
result provides valuable insight.

\begin{lemma} \label{lem2}
For all positive integers $n$, $\displaystyle{[A^n,B]=\sum_{i=1}^n A^{n-i}CA^{i-1}.}$
\end{lemma}
\proof{
The result follows from induction, noting that $n=1$ recovers the relation $[A,B]=C.$ 
}

Alternatively, the result of Lemma \ref{lem2} may be expressed as a sum of symmetric terms as 
$$
[A^n,B] = \frac{1}{2} \sum_{i=1}^n \left( A^{n-i}CA^{i-1} + A^{i-1}CA^{n-i} \right).
$$
Without loss of generality, we may focus only on the expressions $A^{n-i}CA^{i-1} +
A^{i-1}CA^{n-i}$ for which $n-i\geq i-1$. 
Defining
\begin{align}
P(m,\ell) &= A^mCA^\ell + A^\ell CA^m,\label{defP}\\
Q(m,\ell) &= A^mBA^\ell - A^\ell BA^m, \label{defQ}
\end{align}
and noting that
\begin{align*}
P(m,\ell) &= P(\ell,m),\\
Q(m,\ell) &= -Q(\ell,m),
\end{align*}
and
\begin{align*}
P(m,0) &= \{ A^m, C \},\\
Q(m,0) &= [A^m,B ],
\end{align*}
we may then write the result of Lemma \ref{lem2} as
$$
Q(n,0) = \frac12 \sum_{i=1}^n P(n-i,i-1),
$$
or alternatively as
\begin{align}
Q(2k,0) &=\sum_{i=1}^k P(2k-i,i-1),\label{intrel1}\\
Q(2k+1,0) &= \sum_{i=1}^k P(2k+1-i,i-1) + \frac12P(k,k).\label{intrel2}
\end{align}
In this notation, using the commutation relations it is straightforward to verify that
\begin{align}
P(m,\ell) &= P(m+1,\ell-1) - \delta Q(m,\ell-1) - \beta Q(m+1,\ell-1) - \beta Q(m,\ell),
\label{rr1}\\
Q(m,\ell) &= Q(m+1,\ell-1) - P(m,\ell-1), \label{rr2}
\end{align}
leading to a system of two variable recurrence relations.
Further substituting equation (\ref{rr2}) into equation (\ref{rr1}) then gives 
\begin{align}
P(m,\ell) &= P(m+1,\ell-1) - \delta Q(m,\ell-1) - 2\beta Q(m+1,\ell-1) + \beta P(m,\ell-1),
\label{rr1d}\\
Q(m,\ell) &= Q(m+1,\ell-1) - P(m,\ell-1), \label{rr2d}
\end{align}
the significance being that equations (\ref{rr1d}) and (\ref{rr2d}) reduce the second
variable by one on each iteration. This inspires the following lemma.

\begin{lemma} \label{lem3}
Let $x_{i,j}$ and $y_{i,j}$ be numbers satisfying the system of recurrence relations
\begin{align*}
x_{i,j} &= x_{i-1,j-1}+\beta x_{i,j-1}+y_{i,j-1},\\
y_{i,j} &= \delta x_{i,j-1}+2\beta x_{i-1,j-1}+y_{i-1,j-1},
\end{align*}
with 
$$
x_{0,1} = \beta,\ \ x_{1,1} = 1,\ \ y_{0,1} = \delta,\ \ y_{1,1} = 2\beta,
$$
and where we adopt the convention
$$
x_{-1,j} = 0 = y_{-1,j},\ \ x_{j+1,j} = 0 = y_{j+1,j},
$$
for any $j\geq 1$.
Then for $m\geq \ell\geq 1$, we have
$$
A^mCA^\ell + A^\ell CA^m = \sum_{i=0}^\ell x_{i,\ell}\{A^{m+i},C\} - \sum_{n=0}^\ell
y_{n,\ell} [A^{m+n},B].
$$
\end{lemma}
\proof{
Exploiting the notation of equations (\ref{defP}) and (\ref{defQ}), we first use induction to prove the result
\begin{equation}
P(m,\ell) = \sum_{i=0}^jx_{i,j}P(m+i,\ell-j)-\sum_{k=0}^j y_{k,j} Q(m+k,\ell-j)
\label{propequ1}
\end{equation}
for any $j=1,2,\ldots,\ell$. Clearly this holds for $j=1$, recovering equation
(\ref{rr1d}). The induction hypothesis gives
\begin{align*}
P(m,\ell) &= \sum_{i=0}^jx_{i,j}P(m+i,\ell-j)-\sum_{k=0}^j y_{i,j} Q(m+k,\ell-j) \\
&= \sum_{i=0}^jx_{i,j}\left( P(m+i+1,\ell-j-1) + \beta P(m+i,\ell-j-1) - \delta
Q(m+i,\ell-j-1)\right. \\
&\qquad\qquad - \left.2\beta Q(m+i+1,\ell-j-1)\right)\\
& \ \ - \sum_{k=0}^j y_{k,j}\left( Q(m+k+1,\ell-j-1) - P(m+k,\ell-j-1) \right)\\
&= \sum_{i=0}^j\left( x_{i-1,j}+\beta x_{i,j} + y_{i,j} \right) P(m+i,\ell-j-1) +
x_{j,j}P(m+j+1,\ell-j-1) \\
& \ \ - \sum_{k=0}^j\left( \delta x_{k,j}+2\beta x_{k-1,j} + y_{k-1,j}
\right)Q(m+k,\ell-j-1) \\
&\ \ - (2\beta x_{j,j} + y_{j,j})Q(m+j+1,\ell-j-1)\\
&= \sum_{i=0}^{j+1}\left( x_{i-1,j}+\beta x_{i,j} + y_{i,j} \right) P(m+i,\ell-j-1) \\
&\ \ - \sum_{k=0}^{j+1}\left( \delta x_{k,j} + 2\beta x_{k-1,j} + y_{k-1,j} \right)
Q(m+k,\ell-j-1) \\
&= \sum_{i=0}^{j+1} x_{i,j+1} P(m+i,\ell-j-1)- \sum_{k=0}^{j+1}y_{k,j+1}Q(m+k,\ell-j-1),
\end{align*}
where we have made use of the conventions highlighted in the statement of the
Lemma. This proves the identity (\ref{propequ1}). Setting $j=\ell$ as a special
case then proves the Lemma.
}

In the context of Lemma \ref{lem2}, particularly the form given in equations
(\ref{intrel1}) and (\ref{intrel2}), the result of Lemma \ref{lem3} implies that
\begin{align}
Q(2k,0) &= \sum_{i=1}^k\sum_{j=0}^{i-1}\left( x_{j,i-1}P(2k-i+j,0) - y_{j,i-1}Q(2k-i+j,0)
\right), \label{Qeven}\\
Q(2k+1,0) &= \sum_{i=1}^k\sum_{j=0}^{i-1}\left( x_{j,i-1}P(2k+1-i+j,0) -
y_{j,i-1}Q(2k+1-i+j,0) \right)\nonumber\\
&\ \ +\frac12\sum_{i=0}^k\left( x_{i,k}P(k+i,0) - y_{i,k}Q(k+i,0) \right). \label{Qodd}
\end{align}
In other words, we can see that the $Q(n,0)$ reduces to a linear combination of $P(u,0)$
and $Q(v,0)$ terms, in particular where $v$ is less than $n$. Using this idea recursively
gives rise to the following consequence of Lemma \ref{lem3}.

\begin{cor} \label{cor2}
For any $j=1,2,\ldots$, we have
$$
[A^{j},B] = \sum_{k=0}^{j-1}s_k^{(j)}\{ A^k,C\}.
$$
\end{cor}
Making use of the result of Corollary \ref{cor2} allows us to express equation
(\ref{Qeven}) as
\begin{align}
Q(2n,0) &= \sum_{i=1}^n\sum_{m=0}^{2n-1} T^{2n-i,i-1}_m P(m,0), \label{Qeven2}
\end{align}
where
\begin{align}
T^{j-i,i-1}_m = 
\left\{
\begin{array}{l}
\displaystyle{-\sum_{k=0}^{i-1}y_{k,i-1}s^{(j-i+k)}_m, }\ \ \   m=0,1,\ldots,j-i-1,\\
\ \\
\displaystyle{ x_{m-j+i,i-1}-\sum_{k=m-j+i+1}^{i-1}y_{k,i-1}s^{(j-i+k)}_m, } \ \ \  m =
j-i,j-i+1,\ldots,j-2,\\
\ \\
x_{i-1,i-1},\ \ \   m=j-1, 
\end{array}
\right.
\label{defT}
\end{align}
for $i=2,3,\ldots,n$, and $T^{2n-1,0}_m = \delta^{2n-1}_m$ (i.e. the Kronecker delta).
Interchanging the order of summation in (\ref{Qeven2}) trivially gives 
$$
Q(2n,0) = \sum_{m=0}^{2n-1}\sum_{i=1}^n T^{2n-i,i-1}_mP(m,0).
$$
Corollary \ref{cor2}, however, implies 
$$
Q(2n,0)= \sum_{m=0}^{2n-1} s^{(2n)}_mP(m,0),
$$
from which we conclude that
\begin{align}
s^{(2n)}_m = \sum_{i=1}^nT^{2n-i,i-1}_m. \label{Sequ1}
\end{align}
Regarding (\ref{Qodd}), in a similar way we may use Corollary \ref{cor2} to write
\begin{align}
Q(2n+1,0) = \sum_{m=0}^{2n} \left( \sum_{i=1}^nT^{2n+1-i,i-1}_m + \frac12 T^{n,n}_m \right)
P(m,0), \label{Qodd2}
\end{align}
from which we conclude that 
\begin{align}
s^{(2n+1)}_m = \sum_{i=1}^nT^{2n+1-i,i-1}_m + \frac12 T^{n,n}_m. \label{Sequ2}
\end{align}

In order to actually determine the $s^{(k)}_m$ coefficients introduced in Corollary
\ref{cor2}, we first need to find the $x_{i,j}$ and $y_{i,j}$ of Lemma \ref{lem3}, and
then treat equations (\ref{Sequ1}) and (\ref{Sequ2}) as a system of recurrence relations
for the $s^{(k)}_m$. We have as an ``initial condition'' to this system 
$$
s^{(1)}_0 = \frac12,
$$
which arises from the definition of $C$ in equation (\ref{genl1v1q}). For a given
degree $M$ of the polynomial algebra, these equations are solved recursively for
$s^{(k)}_m$.

We are now in a position to give more detail about the structure constants satisfying constraint
(\ref{qjaccon2}). Having (in principle) determined the $s^{(k)}_m$, we use
the result of Corollary \ref{cor2} to express equation (\ref{qjaccon2}) as 
$$
\eta C + \sum_{i=1}^L\omega_i\{A^i,C\} +
\sum_{k=1}^{L+1}\sum_{i=0}^{k-1}\alpha_ks^{(k)}_i\{A^i,C\}=0,
$$
where we have set $\displaystyle{L=\left\lfloor \frac{M}{2} \right\rfloor.}$
Furthermore, for convenience we set $\omega_0=\eta/2$ and interchange the order of the double sum, so
that the constraint becomes
$$
\sum_{i=0}^L\omega_i\{ A^i,C\} +
\sum_{i=0}^L\sum_{k=i+1}^{L+1}\alpha_ks^{(k)}_i\{A^i,C\}=0.
$$
Equating the coefficients of the linearly independent $\{A^i,C\}$ then gives the following
result.

\begin{prop} \label{prop5}
The constraint (\ref{qjaccon2}) is satisfied by
$$
\omega_i = -\sum_{k=i+1}^{L+1}\alpha_k s^{(k)}_i,
$$
for $i=0,1,\ldots, L,$ where $\displaystyle{L=\left\lfloor \frac{M}{2} \right\rfloor}$
and $\displaystyle{\omega_0=\frac{\eta}{2}}$.
\end{prop}

\subsubsection{Summary} \label{algorithm}

We now summarise the algorithm that we use to determine the structure constants of the
degree $M$ polynomial Lie algebra.

\begin{itemize}
\item[Step 1.] Find $x_{i,j}$ and $y_{i,j}$ satisfying the recurrence relations of Lemma
\ref{lem3}.
\item[Step 2.] Using the result of Step 1, determine the coefficients $s^{(j)}_k$ of
Corollary \ref{cor2} by solving the recurrence relations given by equations (\ref{Sequ1}) and
(\ref{Sequ2}).
\item[Step 3.] Using the result of Step 2, give the explicit expressions for the structure
constants $\eta$ and the $\omega_i$ in terms of the $\alpha_i$, $\beta$ and $\delta$ via
Proposition \ref{prop5}.
\end{itemize}

We remark that while we have not given closed form solutions for the structure constants
for arbitrary $M$, we are able to find the structure constants using the recursive
algorithm outlined in Steps 1,2 and 3 above for a {\em specified} value of $M$. The
results of our methods agree with the low order cases presented in 
\cite{Das1,Mar2,Mar3}.

\subsection{Casimir operator for the Lie algebra case}

By analogy with the classical case of Section \ref{casimiropsec}, we consider a Casimir
operator $K$ defined by
\begin{equation}
[A,K] = 0 = [B,K].\label{cascond}
\end{equation}
As in the previous section, we set
$$
L = \left\lfloor \frac{M}{2} \right\rfloor. 
$$
From the low order polynomial algebras studied in \cite{Das1,Mar2,Mar3} and the classical
case we consider the Casimir operator to be of the form
\begin{equation}
K=C^{2}+\sum_{i=1}^{L+1}\frac{1}{2}m_{i}\{A^{i},B\} +
\frac{n}{2}\{A,B^{2}\}+\sum_{i=1}^{M+1}k_{i}A^{i}+\ell_{1}B + \ell_2 B^2. \label{qcasimir}
\end{equation}
We now seek constraints on the coefficients by imposing the two conditions stated above in
(\ref{cascond}). Firstly, after some straightforward manipulation involving the Lie
bracket given in equations (\ref{genl1v1q}) - (\ref{genl3v1q}) of Proposition \ref{prop4}, we have
\begin{align*}
[A,K] &= \sum_{i=1}^{L+1}\alpha_i\{A^i,C\} +
\delta\{B,C\}+2\epsilon C + \beta\{\{A,B\},C\} \\
&\ \  + \sum_{i=1}^{L+1}\frac12 m_i\{A^i,C\} +
\frac{n}{2}\{A,\{B,C\}\} + \ell_1 C + \ell_2\{B,C\}.
\end{align*}
Furthermore, one may verify that
\begin{align*}
\{A,\{B,C\}\} &=  \{\{A,B\},C\} + [[A,C],B]\\
&= \{ \{ A,B\},C\} + \sum_{i=1}^{L+1}\alpha_i[A^i,B]+\beta\{B,C\},
\end{align*}
which then gives
\begin{align*}
[A,K] &= \sum_{i=1}^{L+1}\alpha_i\{A^i,C\} +
\delta\{B,C\}+2\epsilon C + \beta\{\{A,B\},C\} \\
&\ \  + \sum_{i=1}^{L+1}\frac12 m_i\{A^i,C\} +
\frac{n}{2}\{\{A,B\},C\}\} + \sum_{i=1}^{L+1}\frac{n\alpha_i}{2}[A^i,B] +
\frac{n\beta}{2}\{B,C\}+ \ell_1 C + \ell_2\{B,C\}.
\end{align*}
Setting the coefficients of $\{\{A,B\},C\}$ and $\{B,C\}$ to zero then gives
\begin{align*}
n=-2\beta,\ \ \ell_2 = \beta^2-\delta
\end{align*}
respectively, and leaves us with
\begin{align}
[A,K] &= \sum_{i=1}^{L+1}\alpha_i\{A^i,C\} + 2\epsilon C 
+ \sum_{i=1}^{L+1}\frac12 m_i\{A^i,C\} -
\sum_{i=1}^{L+1}{\beta\alpha_i}[A^i,B] +
\ell_1 C.
\label{cas1}
\end{align}
The constraint (\ref{qjaccon2}) from Proposition \ref{prop4} implies that we can
make the substitution
$$
-\sum_{i=1}^{L+1}\alpha_{i}[A^{i},B]
=\eta C+\sum_{i=1}^{L}\omega_{i}\{A^{i},C\}
$$
in (\ref{cas1}), leading to
\begin{align*}
[A,K] &= \sum_{i=1}^{L+1}\alpha_i\{A^i,C\} + 2\epsilon C 
+ \sum_{i=1}^{L+1}\frac12 m_i\{A^i,C\} +
\beta\eta C + \sum_{i=1}^{L}\beta\omega_{i}\{A^{i},C\}
+\ell_1 C.
\end{align*}
Therefore the Casimir property $[A,K]=0$ can be obtained by setting the coefficients of $C$ and
$\{A^i,C\}$ to zero. Namely,
\begin{align*}
\ell_1 &= -2\epsilon-\beta\eta,\\
m_i &= -2\alpha_i-2\beta\omega_i,\ \ i=1,\ldots,L\\
m_{L+1} &= -2\alpha_{L+1}.
\end{align*}
The form of the Casimir operator $K$ is then
\begin{align}
K &= C^2 - \sum_{i=1}^{L+1}\alpha_i\{A^i,B\} - \sum_{i=1}^L\beta\omega_i\{A^i,B\} -
\beta\{A,B^2\}
+\sum_{i=1}^{M+1}k_iA^i - (2\epsilon+\beta\eta) B +(\beta^2-\delta)B^2.
\label{CasimirForm}
\end{align}
Now there remain constraints to be determined by applying $[B,K]=0.$
We have
\begin{align*}
[B,K] &= [B,C^2]  - \sum_{i=1}^{L+1}\alpha_i[B,\{A^i,B\}] -
\sum_{i=1}^L\beta\omega_i[B,\{A^i,B\}] - \beta[B,\{A,B^2\}]+\sum_{i=1}^{M+1}k_i[B,A^i].
\end{align*}
We make use of the easily established relations
\begin{align*}
[B,C^2] &= \sum_{i=1}^M\lambda_i\{A^i,C\} - \beta\{B^2,C\}+\eta\{ B,C\} +
\sum_{i=1}^L\omega_i\{\{A^i,B\},C\} + 2\zeta C,\\
[B,\{A^i,B\}] &= -\{ [A^i,B],B\},\\
[B,\{A,B^2\}] &= -\{B^2,C\}, 
\end{align*}
along with
$$
\{\{ A^i,B\},C\} = \{\{ A^i,C\},B\} - \beta\{ [A^i,B],B\} + \eta[A^i,B] + \sum_{j=1}^L\omega_j[A^i,\{A^j,B\}],
$$
to simplify this expression. That is,
\begin{align}
[B,K] &= \sum_{i=1}^M\lambda_i\{A^i,C\} + \eta\{B,C\} + \sum_{i=1}^L\omega\{\{A^i,C\},B\} +
\sum_{i=1}^L\omega_i\eta[A^i,B]\nonumber\\
&\ \ +\sum_{i=1}^L\sum_{j=1}^L\omega_i\omega_j[A^i,\{A^j,B\}] +
2\zeta C + \sum_{i=1}^{L+1}\alpha_i\{[A^i,B],B\} - \sum_{i=1}^{M+1}k_i[A^i,B]\nonumber\\
&= \sum_{i=1}^M\lambda_i\{A^i,C\} + \sum_{i=1}^L\omega_i\eta[A^i,B]
+\sum_{i=1}^L\sum_{j=1}^L\omega_i\omega_j[A^i,\{A^j,B\}] +2\zeta C \nonumber\\
& \ \  - \sum_{i=1}^{M+1}k_i[A^i,B] 
+ \left\{B,\eta C+\sum_{i=1}^L\omega_i\{A^i,C\} + \sum_{i=1}^{L+1}\alpha_i[A^i,B]\right\}.
\label{Bcalc1}
\end{align}
Clearly the last anti-commutator term in the above expression vanishes due to constraint
(\ref{qjaccon2}). We also have
$$
[A^i,\{A^j,B\}] = [A^{i+j},B] + A^iBA^j - A^jBA^i.
$$
The following result is helpful in further simplifying these expressions.
\begin{lemma} \label{prop6}
Let $\bar{x}_{i,j}$ and $\bar{y}_{i,j}$ be numbers satisfying the system of recurrence relations
\begin{align*}
\bar{x}_{i,j} &= \bar{x}_{i-1,j-1}+\beta \bar{x}_{i,j-1}+\bar{y}_{i,j-1},\\
\bar{y}_{i,j} &= \delta \bar{x}_{i,j-1}+2\beta \bar{x}_{i-1,j-1}+\bar{y}_{i-1,j-1},
\end{align*}
with 
$$
\bar{y}_{0,1} = 0,\ \ \bar{y}_{1,1} = 1,\ \ \bar{x}_{0,1} = 1,
$$
and where we adopt the convention
$$
\bar{y}_{-1,j} = 0 = \bar{x}_{-1,j},\ \ \bar{x}_{j,j} = 0 = \bar{x}_{j+1,j},
$$
for any $j\geq 1$.
Then for $i> j\geq 1$, 
\begin{equation}
A^iBA^j - A^j BA^i = \sum_{k=0}^j \bar{y}_{k,j}[A^{i+k},B] - \sum_{k=0}^{j-1}
\bar{x}_{k,j} \{A^{i+k},C\}. \label{uvrec}
\end{equation}
\end{lemma}
\proof{
Using the notation of equations (\ref{defP}) and (\ref{defQ}), the result we seek to
prove is expressed as
$$
Q(i,j) = \sum_{k=0}^j \bar{y}_{k,j}Q(i+k,0) - \sum_{k=0}^{j-1}
\bar{x}_{k,j} P(i+k,0).
$$
The details of the proof follow the same lines as that of Lemma \ref{lem3}.
}

We remark that the system of recurrence relations are the same as those occuring in
Lemma \ref{lem3}, but with different boundary conditions.
 
Substituting the result of Corollary \ref{cor2} into equation (\ref{uvrec}) gives the
following.
\begin{cor} \label{cor3}
For $i, j\geq 1$,
$$
A^iBA^j - A^jBA^i = \sum_{k=0}^{i+j-1} W_k^{i,j}\{A^k,C\},
$$
where
$$
W_k^{i,j} = \left\{ \begin{array}{l} \displaystyle{\sum_{m=0}^j \bar{y}_{m,j}s_k^{(i+m)}},
\ \ \ k=1,2,\ldots,i-1\\
\ \\
\displaystyle{\sum_{m=k-i+1}^j \bar{y}_{m,j}s_k^{(i+m)} - \bar{x}_{k-i,j}},\ \ \  k=i,i+1,\ldots i+j-1.  \end{array}  \right.
$$
\end{cor}
One may verify that $W^{i,i}_k=0$ for all $k$ as expected.

The outcome of Corollary \ref{cor3} is that we may write
\begin{align*}
[A^i,\{A^j,B\}] &= [A^{i+j},B] + A^iBA^j - A^jBA^i\\
&= \sum_{k=0}^{i+j-1}\left(s_k^{(i+j)} + W^{i,j}_k\right) \{A^k,C\}.
\end{align*}
Setting $\lambda_0 = \zeta$, equation (\ref{Bcalc1}) then becomes
\begin{align*}
[B,K] &= \sum_{k=0}^M\lambda_k\{A^k,C\} + \sum_{i=1}^L\sum_{k=0}^{i-1}\omega_i\eta s_k^{(i)}\{A^k,C\}
+ \sum_{i=1}^{L+1}\sum_{k=0}^{i-1}\alpha_is^{(i)}_k\{A^k,C\}
\\
& \  \ + \sum_{i=1}^L\sum_{j=1}^L\sum_{k=0}^{i+j-1} \omega_i\omega_j\left( s^{(i+j)}_k +
W^{i,j}_k \right) \{A^k,C\}  
- \sum_{i=1}^{M+1}\sum_{k=0}^{i-1}k_is^{(i)}_k\{ A^k,C \}.
\end{align*}
Interchanging the order of all the multiple summations so that the sums over $k$ are on
the outside leads to
\begin{align*}
[B,K] &= \sum_{k=0}^M\lambda_k\{A^k,C\} + \sum_{k=0}^{L-1}F_k\{A^k,C\} + \sum_{k=0}^L
G_k\{A^k,C\} + \sum_{k=0}^{2L-1}Z_k\{A^k,C\}  - \sum_{k=0}^MH_k\{A^k,C\},
\end{align*}
where
\begin{align*}
F_k = \sum_{i=k+1}^L\omega_i\eta s^{(i)}_k,\ \
G_k = \sum_{i=k+1}^{L+1}\alpha_i s^{(i)}_k,
\end{align*}
\begin{align}
H_k = \sum_{i=k+1}^{M+1}k_i s^{(i)}_k, \label{kishere}
\end{align}
and 
$$
Z_k = \left\{ 
\begin{array}{l}
\displaystyle{\sum_{i=1}^L\sum_{j=1}^L\omega_i\omega_j\left( s^{(i+j)}_k + W^{i,j}_k
\right),} \ \ \  k=0,1,\\
\ \\
\displaystyle{\sum_{i=k}^L\sum_{j=1}^L\omega_i\omega_j\left( s^{(i+j)}_k + W^{i,j}_k\right) + \sum_{i=1}^{k-1}\sum_{j=k-i+1}^L\omega_i\omega_j\left( s^{(i+j)}_k + W^{i,j}_k
\right),} \ \ \  k=2,3,\ldots,L,\\
\ \\
\displaystyle{\sum_{i=k-L+1}^L\sum_{j=k-i+1}^L \omega_i\omega_j\left( s^{(i+j)}_k + W^{i,j}_k
\right),} \ \ \  k=L+1,\ldots, 2L-1.
\end{array}
\right.
$$
Setting $[B,K]=0$ and equating coefficients of the linearly independent $\{A^k,C\}$ then
leads to the system of linear equations (keeping in mind $\lambda_0=\zeta$)
\begin{align}
\lambda_i+F_i+G_i+Z_i &= H_i,\ \ i=0,1,\ldots,L-1, \label{Bequ1}\\
\lambda_L + G_L+Z_L &= H_L, \label{Bequ2}\\
\lambda_i + Z_i &= H_i,\ \ i=L+1,\ldots,2L-1, \label{Bequ3}\\
\lambda_i &= H_i,\ \ i=2L,M. \label{Bequ4}
\end{align}
Since there are clearly $L$ equations in (\ref{Bequ1}), 1 equation in (\ref{Bequ2}), $L-1$
equations in (\ref{Bequ3}) and $M-2L+1$ ($= 1$ if $M$ is even and 2 otherwise) equations
in (\ref{Bequ4}), there are $M+1$ linear equations that can be solved for the coefficients
$k_1,k_2,\ldots,k_{M+1}$. We have therefore shown the following.
 
\begin{prop} \label{prop7}
The Casimir operator $K$ is of the form given in (\ref{CasimirForm}), with
coefficients $k_i$ determined by first solving the system
of linear equations (\ref{Bequ1}) - (\ref{Bequ4}) for the $H_k$, and then solving
(\ref{kishere}).
\end{prop}

\subsection{Oscillator Realization}

Let us consider the following deformed oscillator algebra:
\begin{equation}
[N,b^{\dagger}]=b^{\dagger},\quad [N,b]=-b, \quad
b^{\dagger}b=\Phi(N),\quad bb^{\dagger}=\Phi(N+1)
\label{defoscalg}
\end{equation}
and realizations of the form
\begin{equation}
A=A(N),\quad B=b(N)+b^{\dagger}\rho(N)+\rho(N)b.
\label{qreal}
\end{equation}
Defining
$$
\Delta A(N) = A(N+1)-A(N),
$$
by analogy with Lemma \ref{lem1} in the classical case, we have the easily 
established relations
\begin{align*}
[A(N),b^{\dagger}] &= b^{\dagger}\Delta A(N),\\
[A(N),b] &= -\Delta A(N)b,\\
\{A(N),b^{\dagger}\}& = b^{\dagger}(A(N+1)+A(N)),\\
\{A(N),b\} & = (A(N+1)+A(N))b.
\end{align*}
We may obtain the realization of the generator $C$ by applying the first
relation~\eqref{genl1v1q} of the polynomial algebra, giving
\begin{equation}
C = [A,B]=b^{\dagger}\  \Delta A(N)\ \rho(N)-\rho(N)\ \Delta A(N)\ b.
\label{qreal2}
\end{equation}
The second relation~\eqref{genl2v1q} provides two equations to constrain $A(N)$ and $b(N)$
as in the classical case. Applying the realization, one has
\begin{align*}
[A,C]& =b^{\dagger} \left(\Delta A(N)\right)^2 \rho(N)+\rho(N) \left(\Delta
A(N)\right)^{2} b\\
&= \sum_{i=1}^{L+1}\alpha_{i}(A(N))^i+\delta (b(N)+b^{\dagger}\rho(N)+\rho(N)b)+\epsilon
\\
&\ \ +\beta
\left( 2A(N) b(N)+ b^\dagger(A(N+1)+A(N)) \rho(N) + \rho(N)(A(N+1)+A(N))b
\right) .
\end{align*}
This gives two functional equations for the functions $A(N)$ and $b(N)$:
\begin{align}
(\Delta A(N))^2 &=\delta +\beta (A(N)+A(N+1)), \label{qsolcAB1} \\
0 &= \sum_{i=1}^{L+1}\alpha_{i}A(N)^{i}+\delta b(N)+\epsilon+2\beta A(N) b(N). \label{qsolcAB2}
\end{align}
As in the classical case, the conditions (\ref{qsolcAB1}) and
(\ref{qsolcAB2}) can be solved for $A(N)$ and $b(N)$, for two distinct cases. 
We have
\begin{align}
A(N) &= \left\{ \begin{array}{rl}
\displaystyle{\sqrt{\delta}N + c_1},& \beta=0,\\
\displaystyle{-\frac{\beta}{8}-\frac{\delta}{2\beta}+\frac{\beta}{2}(N+c_{1})^{2}},& \beta\neq 0,
\end{array} \right.
\label{Asol}
\end{align}
with $c_1$ an arbitrary constant,
\begin{align}
b(N) &=  -\frac{4}{4A'(N)^2-\beta^2}\sum_{i=1}^{L+1}
\alpha_{i}A(N)^i+\epsilon,
\label{Bsol}
\end{align}
where $A'$ denotes the usual derivate of the function $A$.

At this stage we point out that equation (\ref{qjaccon2}), which gives a relationship
between the structure constants of the Lie algebra, also has a corresponding expression in terms of the
realization. This expression, which actually arises as the coefficient of both $b$ and $b^\dagger$ 
upon substitution of the realization into (\ref{qjaccon2}), is given by
\begin{align}
\eta\Delta A(N) + \sum_{i=1}^L\omega_i\Delta A(N)\left( A(N)^i + A(N+1)^i
\right)+\sum_{i=1}^{L+1}\alpha_i\left( A(N+1)^i - A(N)^i \right)=0.
\label{jacobireal}
\end{align}

Now applying the realization to the third relation~\eqref{genl3v1q} of the Lie algebra leads to
\begin{align*}
[B,C]&=\left( b^{\dagger}\right)^2\rho(N)\rho(N+1)(\Delta A(N)-\Delta
A(N+1))+\rho(N)\rho(N+1)(\Delta A(N)-\Delta A(N+1))b^2 \\
& \ \  +b^{\dagger} \rho(N) \Delta A(N)(b(N+1)-b(N))+ \rho(N) \Delta A(N)(b(N+1)-b(N))b \\
& \ \ -2\Phi(N)\rho(N-1)^2\Delta A(N-1)+2\Phi(N+1)\rho(N)^2 \Delta A(N) \\
& = \zeta + \sum_{i=1}^M\lambda_{i}A(N)^i +\eta \left(b(N)+b^{\dagger}\rho(N)+\rho(N)b\right) \\
& \ \ -\beta \left( \left(b^\dagger\right)^2 \rho(N)\rho(N+1) +
\rho(N)\rho(N+1)b^2\right)\\
&\ \  -\beta\left(
b^\dagger (b(N+1)+b(N))\rho(N) + \rho(N)(b(N+1)+b(N))b\right) \\
&\ \ -\beta\left(  b(N)^2+\rho(N-1)^2\Phi(N) + \rho(N)^2\Phi(N+1) \right) \\
&\ \ + \sum_{j=1}^L\omega_{i} \left( b^\dagger\left( A(N+1)^j + A(N)^j \right) \rho(N) +
\rho(N)\left( A(N+1)^j+A(N)^j \right)b + 2A(N)^jb(N) \right) .
\end{align*}
This gives rise to the three constraint relations
\begin{align}
& \Delta A(N)-\Delta A(N+1)=-\beta  \label{qeq3c1} \\
& \Delta A(N)(b(N+1)-b(N))=-\beta( b(N+1)+b(N))+\sum_{i=1}^L\omega_{i}(A(N+1)^i+A(N)^i)+\eta
\label{qeq3c2} \\
& -2 \Phi(N)\rho(N-1)^2\Delta A(N-1) +2 \Phi(N+1)\rho(N)^2\Delta A(N) \nonumber \\
& \ \ =\zeta+\sum_{i=1}^M\lambda_{i} A(N)^i-\beta b(N)^2 
-\beta (\rho(N-1)^2\Phi(N)+\rho(N)^2\Phi(N+1))+\eta b(N)\nonumber\\
&\ \ \  \ +\sum_{j=1}^L2\omega_{j} A(N)^{j}b(N).\label{qeq3c3}
\end{align}
The first two relations, however, can be seen to be redundant given (\ref{qsolcAB1}),
(\ref{qsolcAB2}) and (\ref{jacobireal}). This can be seen from the following calculation.

In the first instance, 
let us consider~\eqref{qsolcAB1} at $N$ and $N+1$ and substract the two equations. This
leads directly to equation (\ref{qeq3c1}).

Now let us multiply the equation~\eqref{qeq3c2} by $\Delta A(N)$ and replace the expression
for $(\Delta A(N))^2$ from~\eqref{qsolcAB1}. We obtain
\begin{align}
& (\delta +\beta A(N)+A(N+1))(b(N+1)-b(N)) \nonumber\\
&\ \ \ \ =-\beta \Delta A(N)(b(N+1)+b(N))
+\Delta A(N) \sum_{i=1}^L \omega_{i} (A(N+1)^i+A(N)^i)+\Delta A(N) \eta
\label{interim1}
\end{align}
We use the equation~\eqref{qsolcAB2} at $N+1$ and $N$ and take their difference, giving
the relation
\begin{align}
\delta( b(N+1)-b(N)) = \sum_{i=1}^{L+1}\alpha_i\left( A(N)^i - A(N+1)^i
\right)+2\beta(A(N)b(N) - A(N+1)b(N+1)).
\label{interim2}
\end{align}
Substituting relation (\ref{interim2}) in for the expression $\delta(b(N)+1)-b(N))$ in
equation (\ref{interim1}) then gives precisely the condition (\ref{jacobireal}).
Thus equation (\ref{jacobireal}) (which we recall arises from the Jacobi identity) tells
us that~\eqref{qeq3c1} and~\eqref{qeq3c2} do not provide further constraints. 
We do, however, remain with equation~\eqref{qeq3c3} as in the classical case. We summarise
our results on the realization of the polynomial Lie algebra in the following Proposition, 
which serves as the quantum analogue of Proposition \ref{realthm}.
\begin{prop} \label{qrealprop}
The polynomial Lie algebra ${\cal L}_M$, has realization 
in terms of the deformed oscillator algebra with relations (\ref{defoscalg}) for the
generators $A$, $B$, and $C$ given
by (\ref{qreal}) and (\ref{qreal2}). The functions $A(N)$ and $b(N)$ are given by
(\ref{Asol}) and (\ref{Bsol}) respectively, and the constraint (\ref{qeq3c3}) is satisfied
by $\Phi(N)$ and $\rho(N)$.
\end{prop}

Now we turn to investigate the Casimir operator in the deformed oscillator realization. 
Using the form of the Casimir operator $K$ of equation (\ref{CasimirForm}), we have the following realization for $K$:
\begin{align*}
K &= \left( b^\dagger \right)^2
\left( \Delta A(N+1)\Delta A(N) - \beta(A(N+2)+A(N)) + \beta^2-\delta
\right)\rho(N)\rho(N+1) \\
&\ \  +\left( \Delta A(N+1)\Delta A(N) - \beta(A(N+2)+A(N)) + \beta^2-\delta\right)\rho(N)\rho(N+1) 
b^2\\
& \ \ +b^\dagger 
\left( -\beta(A(N+1) + A(N))(b(N+1)+b(N)) - \sum_{i=1}^{L+1}\alpha_i\left(
A(N+1)^i + A(N)^i\right) \right.\\
&\ \ \quad \quad \left.- \sum_{i=1}^L\beta\omega_i\left( A(N+1)^i + A(N)^i \right) -
2\epsilon - \beta\eta + (\beta^2-\delta)(b(N+1) + b(N))\right)\rho(N)\\
& \ \ + \left( -\beta(A(N+1) + A(N))(b(N+1)+b(N)) - \sum_{i=1}^{L+1}\alpha_i\left(
A(N+1)^i + A(N)^i\right) \right.\\
&\ \ \quad \quad \left. - \sum_{i=1}^L\beta\omega_i\left( A(N+1)^i + A(N)^i \right) -
2\epsilon - \beta\eta + (\beta^2-\delta)(b(N+1) + b(N))\right)\rho(N) b\\
& \ \ 
- \left( \Delta A(N-1) \right)^2\rho(N-1)^2\Phi(N) - \left(\Delta
A(N)\right)^2\rho(N)^2\Phi(N+1) - \sum_{i=1}^{L+1}2\alpha_iA(N)^ib(N) \\
&\ \ - \sum_{i=1}^L2\beta\omega_iA(N)^ib(N) - 2\beta A(N)b(N)^2 - 2\beta A(N)\rho(N-1)^2\Phi(N) -
2\beta A(N)\rho(N)^2\Phi(N+1) \\
&\ \ + \sum_{i=1}^{M+1}k_i A(N)^i - (2\epsilon+\beta\eta)b(N) +
(\beta^2-\delta)\left(b(N)^2 + \rho(N-1)^2\Phi(N) + \rho(N)^2\Phi(N+1)\right).
\end{align*}
The condition that $K$ should not depend explicitly on $b$, $b^\dagger$ or their
powers can be achieved by setting the coefficients of those terms to zero in the above
expression for $K$. This implies the following two constraints, along with the actual form
of $K$:
\begin{align}
0 &= \Delta A(N+1)\Delta A(N) - \beta(A(N+2)+A(N)) + \beta^2-\delta 
\label{qcasimirc1} \\
0 &= -\beta(A(N+1) + A(N))(b(N+1)+b(N)) - \sum_{i=1}^{L+1}\alpha_i\left(
A(N+1)^i + A(N)^i\right)  \nonumber\\
&\ \ - \sum_{i=1}^L\beta\omega_i\left( A(N+1)^i + A(N)^i \right) -
2\epsilon - \beta\eta + (\beta^2-\delta)(b(N+1) + b(N))
\label{qcasimirc2} \\
K &= - \left( \Delta A(N-1) \right)^2\rho(N-1)^2\Phi(N) - \left(\Delta
A(N)\right)^2\rho(N)^2\Phi(N+1) - \sum_{i=1}^{L+1}2\alpha_iA(N)^ib(N) \nonumber \\
&\ \ - \sum_{i=1}^L2\beta\omega_iA(N)^ib(N) - 2\beta A(N)b(N)^2 - 2\beta A(N)\rho(N-1)^2\Phi(N) -
2\beta A(N)\rho(N)^2\Phi(N+1) \nonumber \\
&\ \ + \sum_{i=1}^{M+1}k_i A(N)^i - (2\epsilon+\beta\eta)b(N) +
(\beta^2-\delta)\left(b(N)^2 + \rho(N-1)^2\Phi(N) + \rho(N)^2\Phi(N+1)\right).
\label{qcasimirc3}
\end{align}
It turns out that the constraints (\ref{qcasimirc1}) and (\ref{qcasimirc2}) are also
redundant given the result of Proposition \ref{qrealprop} above. We clarify this in what
follows.

Expressing equation (\ref{qeq3c1}) as
\begin{align}
\Delta A(N) = \Delta A(N+1) - \beta,
\label{interim3}
\end{align}
we substitute (\ref{interim3}) in for $\Delta A(N)$ in equation (\ref{qcasimirc1}), and
then make use of equation (\ref{qsolcAB1}) at $N+1$ to substitute in for 
$(\Delta A(N+1))^2$. The expression then simplifies to equation (\ref{qeq3c1}), which
implies the constraint (\ref{qcasimirc1}) provides no new information.

Finally, taking relation (\ref{qsolcAB2}) in the form
$$
-\sum_{i=1}^{L+1}\alpha_iA(N)^i - \epsilon - \delta b(N) - \beta A(N) b(N) = \beta A(N)
b(N),
$$
and also considering this form at $N+1$, we substitute these expressions into equation
(\ref{qcasimirc2}) which then becomes
$$
-\sum_{i=1}^L\beta\omega_i(A(N+1)^i + A(N)^i)+\beta \Delta A(N)(b(N+1)-b(N)) - \beta\eta +
\beta^2(b(N+1)+b(N))=0.
$$
Multiplying by $\Delta A(N)$ and using (\ref{jacobireal}), (\ref{qsolcAB1}) and
(\ref{qsolcAB2}) at $N$ and $N+1$ shows that the identity is trivially satisfied.
In summary, we have proved the following result.

\begin{prop}
The Casimir operator $K$ of the polynomial Lie algebra ${\cal L}_M$ can be realized in
terms of the deformed oscillator algebra with relations (\ref{defoscalg}). The explicit
expression for $K$ is given by (\ref{qcasimirc3}).
\end{prop}

We remark that in some sense the form of $K$ given in (\ref{qcasimirc3}) can be considered
a further constraint on the functions $\Phi(N)$ and $\rho(N)$. With this form of $K$ in
addition to the constraint (\ref{qeq3c3}), we may determine $\Phi(N)$, with $\rho(N)$
being a polynomial function of $N$.

\subsection{Application to superintegrable systems}

These realizations as defomed oscillator algebras of polynomial associative algebras can be used to
calculate algebraically the energy spectrum and total number of degeneracies per level of quantum superintegrable systems. Before discussing
how these constructions can be applied, let us present some definitions concerning superintegrability
\cite{Mil2}.

\begin{defo}
A classical Hamiltonian system in $n$ dimensions is (polynomially) superintegrable
if it admits $n + k$ (with $k =1,\ldots,n-1$) functionally independent constants of
the motion that are polynomial in the momenta and are globally defined except possibly for
singularities on a lower dimensional manifold. It is minimally (polynomially) superintegrable
if $k = 1$ and maximally (polynomially) superintegrable if $k=n-1$.
\end{defo}

Note that as mentioned in \cite{Mil2}, many distinct $n$-subsets of the $2n - 1$ polynomial constants of the motion for a superintegrable system could be in
involution and in a such case the system would be called multi-integrable.

\begin{defo}
A quantum system in $n$ dimensions is superintegrable (of finite-order) if it admits $n + k$, $k =
1,\ldots,n$ algebraically independent finite-order partial differential operators
$L_{1}=H,\ldots,L_{n}+k$ in the variables $x$ globally defined, such that $[H,L_{j}]=0$. 
Again, it is minimally superintegrable (of finite-order) if $k =1$ and maximally superintegrable (of finite-order) if $k=n-1$.
\end{defo}

A direct consequence of these definitions is that all two-dimensional superintegrable systems are
maximally superintegrable. This means that for a two-dimensional superintegrable system, two
other integrals of motion $A$ and $B$ can commute with the Hamiltonian (i.e. $[H,A]=[H,B]=0$) and
they will generate a non-Abelian algebra. This algebra, if it closes under polynomial relations, will be of
the form studied in Section \ref{Section3} and given in Proposition \ref{prop4}. In such a case, the structure constants will
be polynomial in the Hamiltonian, however as it commutes with generators, this does not modify the results
obtained. The maximal order of these polynomials is bounded by the order of the left side of second
and third relation of polynomial associative algebra. The Casimir can be written only in terms of
the Hamiltonian as a polynomial. The maximal order of this polynomial is constrained by the order of
$K$ is terms of the generators and given by $M+1$.

It was shown that some higher dimensional systems \cite{Das2,Mar1} have a specific structure in
which other integrals ($F_{i}$) are also central elements and form with the Hamiltonian an Abelian
subalgebra i.e.
\begin{equation}
[F_{i},H]=[F_{i},F_{j}]=[F_{i},A]=[F_{i},B]=[F_{i},C]=0.
\end{equation}

These supplementary integrals take for example the form of a monopole charge or angular momentum for
which we know eigenvalues. In such a case, the form of the commutation relations in this paper would
remain valid. The structure constants of the polynomial associative algebra, however, will be
polynomial functions not only of the Hamiltonian but also of these other integrals of motion
($F_{i}$). Thus in the study of the algebra's realization in terms of deformed oscillator algebras and
its representations, we will fix the energy ($H\psi=E\psi$) and these other integrals
($F_{i}\psi=f_{i}\psi$). The Casimir operator $K$ of this quadratic algebra is thus given in terms
of the generators and will be also rewritten as a polynomial of $H$ and $F_{i}$. In this case the
order of these polynomials is bounded by the left side of the second and third relations of the
polynomial associative algebra. However, as these others form an Abelian subalgebra, the form of the
Casimir in terms of generators $A$, $B$ and $C$ is also not affected. We can thus introduce an energy
dependent Fock space of dimension p+1 defined by
\begin{equation*}
H|f_{1},\ldots,f_{n};E,n>=E|f_{1},\ldots,f_{n};E,n>,
\end{equation*}
with
\[ N|f_{1},\ldots,f_{n};E,n>=n|f_{1},\ldots,f_{n};E,n> \]
\[ b|f_{1},\ldots,f_{n};E,0>=0.\]
and the action of operators $b$ and $b^{\dagger}$ are given by
\begin{equation*}
b^{\dagger}|n>=\sqrt{\Phi(f_{1},\ldots,f_{n};E,n+1,u)}|f_{1},\ldots,f_{n};E,n+1>,
\end{equation*}
\[b|n>=\sqrt{\Phi(f_{1},\ldots,f_{n};E,n,u)}|f_{1},\ldots,f_{n};E,n-1>.\]

We furthermore have the existence of finite dimensional unitary representations if we impose the following constraint
\begin{equation*}
\Phi(f_{1},\ldots,f_{n};E,p+1,u)=0, \quad \Phi(f_{1},\ldots,f_{n};E,0,u)=0,\quad \Phi(f_{1},\ldots,f_{n};E,n,u)>0, \quad \forall \; n>0 .
\end{equation*}
The energy $E$ and the constant $u$ can be obtained from this set of constraints that are algebraic
equations. The dimension of the finite-dimensional unitary representations is given by $p+1$.

In the case of three-dimensionnal systems \cite{Das2}, it was shown in an extended Kepler system
that other integrals do not form an Abelian subalgebra and thus do not commute with $A$ and $B$. 
We can, however, identify many structures of quadratic algebra with three generators in which other integrals
can play the role of the Hamiltonian in the structure constant.

Let us point out that this method is very convenient for determining the energy spectrum of various
superintegrable systems. Generally it is not guaranteed that integrals of motion close at a given order and
it was pointed out one needs to sometimes to take further commutators between integrals to generate
integrals that would close.

Furthermore, it was observed that in some cases integrals of motion can close at a given order
but the finite dimensional unitary representations do not provide all the levels and total number of degeneracies
\cite{Mar5}. It was shown, however, that it is still possible to generate higher order polynomial
algebras that provide all the appropriate levels. Through direct but involved calculation, it can be shown
that the degeneracies are correct but for a fixed level given by the union of finite dimensional unitary
representations \cite{Mar6}. This algebraic technique can also generate non-physical solutions that
one needs to remove. We point out that careful analysis should be performed and further study on
the method itself needs to be done.

The polynomial Poisson algebra can be calculated in the context of superintegrable classical systems. However, this is
not clear how to obtain information on the systems from this algebraic structure in the classical context. It can be used in context of the classification of these systems
\cite{Kre1,Mil1}. We can also observe by looking at the classical and quantum analogue how the polynomial Poisson algebra is
deformed into a polynomial associative algebra with higher order correction terms in the Planck
constant.

\section{Examples of Polynomial Lie algebra} \label{Section4}

\subsection{Systems associated with cartesian coordinates}

Let us investigate further the case of the polynomial algebra of arbitrary order related to many examples of 2D superintegrable systems in Euclidean space 
with separation of variables in Cartesian coordinates \cite{Mar4}
$$
[A,B]=C,\quad [A,C]=\delta B,\quad [B,C]=\sum_{i=1}^M\lambda_{i}A^{i}+\zeta.
$$
The Jacobi identity is trivially satisfied since in Proposition \ref{prop4}
$$
\omega_{i}=0,\quad \eta =0,\quad 
\alpha_{i}=0,\quad \beta =0,\quad \epsilon=0.
$$
Constraint (\ref{qeq3c3}) for the Casimir operator reduces to
$$
-2\Phi(N)\rho(N-1)^2\sqrt{\delta} + 2\Phi(N+1)\rho(N)^2\sqrt{\delta} = \zeta + \sum_{i=1}^M\lambda_iA(N)^i.
$$
The solutions for the functions $A(N)$, $b(N)$ are 
\[ A(N)=\sqrt{\delta}N+c_1,\quad b(N)=0  \]
and the Casimir operator is given by 
$$
K=-2\delta\rho(N-1)^2\Phi(N) - 2\delta\rho(N)^2\Phi(N+1) + \sum_{i=1}^{M+1}k_iA(N)^i,
$$
which can be computed explicitly to any given order $M$ using the result of Proposition \ref{prop7} 
(i.e Proposition \ref{prop7} is used to the find the $k_i$ for a fixed order $M$). 
The function $\rho(N)$ can be taken to be a constant in this case.

\subsection{Systems associated with polar coordinates}

Let us describe another class of polynomial algebra of arbitrary order that includes many 2D superintegrable systems in Euclidean space and allowing separation of variables in polar coordinates, such as the TTW systems \cite{Tre1,Kal1}.
$$
[A,B]=C,
$$ 
\[ [A,C]=\alpha_{1} A+\alpha_{2}A^{2}+\delta B+\epsilon +\beta \{A,B\}, \]
\[ [B,C]=\sum_{i=1}^M\lambda_{i}A^{i}-\beta B^{2}+\eta B +\omega_{1}\{A,B\} + \zeta. \]
The constraint equation (\ref{qjaccon2}) of Proposition \ref{prop4} implies
$$
\eta=-\alpha_{1},\quad \omega_{1}=-\alpha_{2}.
$$
The constraints on the parameters of the Casimir operator are
$$
k_{1}=2 \zeta -\alpha_{1}\alpha_{2},\quad k_{2}=\lambda_{1}-\alpha_{2}^{2},
$$
\[ \sum_{i=2}^{M}\lambda_{i}\{C,A^{i}\}=\sum_{i=3}^{M+1}k_{i}[A^{i},B]. \]
This is a system of linear equations in $k_{i}$ that can be solved at any order $M$.
The solution takes the form
\begin{align*}
A(N)&=\frac{\beta}{2}\left( \left((N+c_1)^{2}-\frac{1}{4}\right)-\frac{\delta}{\beta^{2}}\right),
\\
b(N) &=-\frac{\alpha_{2}}{4} \left((N+c_1)^{2}-\frac{1}{4}\right)+\left(\frac{-\beta \alpha_{1}+\alpha_{2}\delta}{2\beta^{2}}\right) \\
& \ \ -\left(\frac{-2\beta \alpha_{1} \delta +\alpha_{2} \delta^{2} +4
\beta^{2}\epsilon}{4\beta^{2}}\right) \frac{1}{   \left((N+c_1)^{2}-\frac{1}{4}\right)}.
\end{align*}
We have two equations to obtain the structure functions, namely
\begin{align*}
& -2 \Phi(N) \rho(N-1)^{2} \Delta A(N-1)^{2} +2 \Phi(N+1) \rho(N)^{2} \Delta A(N)^{2} \\
&\ \ \ \ = \sum_{i=1}^{M} \lambda_{i} A(N)^{i}-\alpha_{1} b(N) +2 \omega_{1} b(N)A(N)
-\beta (\rho^{2}(N-1) \Phi(N)+\rho^{2}(N)\Phi(N+1)),
\end{align*}
along with the expression for $K$ from equation (\ref{qcasimirc3}), which for this case becomes
\begin{align*}
K &= - \left( \Delta A(N-1) \right)^2\rho(N-1)^2\Phi(N) - \left(\Delta
A(N)\right)^2\rho(N)^2\Phi(N+1) - 2\alpha_1A(N)b(N)-2\alpha_2A(N)^2b(n) \nonumber \\
&\ \ +2\beta\alpha_2A(N)b(N) - 2\beta A(N)b(N)^2 - 2\beta A(N)\rho(N-1)^2\Phi(N) -
2\beta A(N)\rho(N)^2\Phi(N+1) \nonumber \\
&\ \ + \sum_{i=1}^{M+1}k_i A(N)^i - (2\epsilon+\beta\eta)b(N) +
(\beta^2-\delta)\left(b(N)^2 + \rho(N-1)^2\Phi(N) + \rho(N)^2\Phi(N+1)\right).
\end{align*}
In particular, an algebraic derivation of the spectrum of the TTW (Tremblay-Turbiner-Winternitz)
Hamiltonian could be obtain using deformed oscillator algebra this is an open problem that would
need to be investigated. As mentioned this paper is devoted to showing the existence of a realization beyond
the quartic case and the construction of the Casimir operator, however it is clear that many examples with integrals of motion of higher order could be studied using
deformed oscillator algebras and results of this paper.

\section{Conclusion}

In this paper, we presented the most general polynomial Lie algebra generated by integrals of order
two and $M$, and presented constraints on the structure constants arising from the Jacobi identity. 
We also constructed the Casimir operator and obtained further constraints on the parameters. 
Explicit formulae could be obtained at an arbitrary, fixed order $M$ and can be applied to a given
example using the algorithm outlined in Section \ref{algorithm}. 
We found, however, that the general solution is difficult and it is clear that further work on this
problem could be undertaken.
We derived many identites concerning various commutators and anti-commutators involving the
generators of the polynomial Lie algebra.
We showed that realizations of the Polynomial Lie algebra via a deformed oscillator algebra exist as
a result of the Jacobi identity.
We also pointed out that in the classical case the most general polynomial Poisson algebra can be
put in the form of the classical analogue of a deformed oscillator algebra when the Jacobi identity is
imposed.
We also explicitly constructed the Casimir operator for the classical case. 

As described in Section \ref{Section3}, the deformed oscillator algebra is a very convenient tool to
study representations and algebraically obtain the energy spectrum of superintegrable systems.
We discussed two particular classes of such polynomial algebras that were observed in the context of
superintegrable systems.

The case of higher dimensional systems would necessitate the study of polynomial algebra with more
than three generators and have investigation of the structure of the subsets of integrals that
commute together. 
The application of realizations via deformed oscillator algebras beyond a class of systems in which other
integrals form an abelian sublgebra with the Hamiltonian is a relatively unexplored subject for the case
of deformed Kepler systems \cite{Das2}. 
This could provide insight to a more systematic study of the algebraic derivation of energy spectrum
of superintegrable systems in more than two dimensions.

Let us point out that quadratic algebras can also be used in the context of systems involving reflection
operators \cite{Gen2}. It is likely that higher order polynomial associative algebras could have
application also in this context and thus formulae that provide realization as oscillator algebras.
We further speculate that contiuned research into the algebraic structures we have introduced and studied in this paper
is likely to reveal applications beyond superintegrable systems in quantum mechanics.

\section*{Acknowledgments} 

The research of I.M. was supported by the Australian Research Council through Discovery
Early Career Researcher Award DE130101067.


\end{document}